\documentclass[aps,twocolumn]{revtex4-1}
\usepackage{graphicx}
\usepackage[justification=justified,width=\linewidth]{caption}
\usepackage{subcaption}
\usepackage{amsmath}
\usepackage{amsfonts}
\usepackage{amsthm}
\usepackage{amssymb}
\usepackage{amsbsy}
\usepackage{wasysym}
\usepackage{bm}
\usepackage{mathrsfs}
\usepackage{color}
\usepackage{times}
\usepackage[resetlabels]{multibib}

\begin{document}
\title{Effect of Annealed Disorder on Phase Separation Kinetics and Aging Phenomena in Fluid Mixtures}
\author {Rounak Bhattacharyya}
\author {Bhaskar Sen Gupta}
\email{bhaskar.sengupta@vit.ac.in}
\affiliation{Department of Physics, School of Advanced Sciences, Vellore Institute of Technology, Vellore, Tamil Nadu - 632014, India}
\date{\today}

\begin{abstract} We use state-of-the-art molecular dynamics simulations to study the effects of annealed disorder on the phase separating kinetics and aging phenomena of a segregating binary fluid mixture. In the presence of disorder, we observe a dramatic slowing down in the phase separation dynamics. The domain growth follows the power-law with a disorder-dependent exponent. Due to the energetically favorable positions, the domain boundary roughens which modifies the correlation function and structure factor to a non-Porod behavior. The correlation function and structure factor provide clear evidence that the superuniversality does not hold in our system. The role of annealed disorder on the non-equilibrium aging dynamics is studied qualitatively by computing the two-time order parameter autocorrelation function. The decay of the correlation function slows down significantly with the disorder. This quantity exhibits scaling laws with respect to the ratio of the domain length at the observation time and the age of the system. We find the scaling laws hold good for the disordered system and therefore, robust and generic to such segregating fluid mixtures. 
\end{abstract}
\maketitle

\section{Introduction} 
Research attention in statistical physics has developed rapidly to understand the out of equilibrium phenomena over the past few decades. In this context, there is a paramount interest to investigate the evolution of a phase separating multi-component mixture with time, due to its application in diverse fields, e.g., magnetic materials, colloidal mixtures, polymer solutions, glasses, metallic alloys, binary liquids \cite{Fisher,Stanley,Binder,Jones,Bray}. When a system of homogeneous multicomponent mixture is rendered thermodynamically unstable by a swift quench inside the miscibility gap, domains with different ordered phases form and expand with time until the system attains a local equilibrium. The domain coarsens due to the organization of matter, mostly dominated by diffusion and can be seen in almost all phase separating systems \cite{Binder1,Siggia,Furukawa,Miguel,Tanaka,Beysens,Tanaka1,Kendon,Puri,Dutt,Laradji,Thakre,Ahmad}.\\
A single time dependent characteristic length $\ell(t)$ usually characterizes the domain growth pattern or domain morphology \cite{Bray}. The same can be obtained by calculating two-point equal time correlation function $\mathrm{C}_{\psi\psi}(\vec{r},t)$, where $\vec{r}$ is the distance between two spatial points and $t$ is the time after quench. So far, it is well demonstrated that pattern formation or coarsening of domains is a scaling phenomenon and exhibits the form ${C}_{\psi\psi}(\vec{r},t)=g[r/\ell(t)]$ \cite{Binder2} where $g(x)$ is the scaling function. The  average domain size $\ell(t)$ follows the power law: $\ell(t) \sim t^{\alpha}$, $\alpha$  defines the growth exponent. The value of the exponent depends on the apposite coarsening mechanism which drives phase separation.

Coarsening mechanism depends on the character of the system. For instance, in solid-solid mixtures (e.g., metallic alloys), diffusion dominates the domain growth whereas hydrodynamic effects contribute significantly in fluid-fluid mixtures. For the earlier the growth exponent $\alpha = 1/3$  which is attributed to Lifshitz-Slyozov (LS) law \cite{Bray,Binder3}. This diffusive regime is very transient in fluid-fluid mixtures and we experience a prompt crossover from diffusive to the hydrodynamic regime. Here  $\alpha$ takes two distinct values, $\alpha=1$ in viscous hydrodynamic regime followed by the inertial hydrodynamic regime with $\alpha=2/3$. The above values of growth exponent are universal and pertinent to pure and isotropic systems \cite{Siggia,Furukawa}. 

Another important characteristic of the non-equilibrium phase separation dynamics is the aging phenomena. The out of equilibrium system changes properties with growing age and it has a fundamental importance in the diverse fields of science and technology. Unlike equilibrium dynamics, the time translational invariance does not hold here. Aging behavior is best characterized by the two-time correlation function $C_{ag}(t,t_w)$ where $t$ and $t_w$ are the observation time and waiting time respectively. In the non-equilibrium state $C_{ag}(t,t_w)$ shows a scaling behavior as $C_{ag}(t/t_w) \sim (\ell/\ell_w)^{-\lambda}$.   Fisher and Huse \cite{Fisher1} introduced the bounds of $\lambda$ for the nonconserved order-parameter dynamics of spin glasses as $\frac{d}{2} \leq \lambda \leq d$. Later a modified lower bound of $\lambda$ was proposed by Yeung, Rao, and Desai \cite{Yeung} for both conservative and non-conservative order-parameter dynamics as $\lambda \geq \frac{\beta + d}{2}$. For the segregating fluid mixtures, the scaling function shows a power-law behavior in the diffusive regime similar to the spin glasses.  However, a crossover from a power-law to exponential behavior is observed in the viscous hydrodynamic regime \cite{sahmed}. This exponential decay is attributed to the fast advective field in the hydrodynamic regime. 

Based on the above discussion we can conclude that the kinetics of phase separation for pure systems is well understood. However, real experimental systems are not free from impurity and this makes the subject more challenging. The disorder strongly influences the kinetics and the domain morphology in a nontrivial manner. Generally, the disorder we find in experimental systems is mainly of two types, (i) quenched or frozen-in or immobile impurities (ii) annealed or mobile impurities.  Quenched disorder acts as a background random potential for fluctuating degrees of freedom. On the other hand, annealed disorder is an additional degree of freedom in the system which is ergodic. In soft matter, disorder impurities can rearrange themselves, and therefore, annealed disorder is more prevalent in soft systems. Numerous studies have been performed so far to understand the diffusion-driven coarsening in Ising systems with quenched disorder \cite{Puri1,Huse,Grest,Srolovitz,Puri2,Bray2,Rao,Paul,Henkel,Aron}. But the phase-separation kinetics with annealed disorder is still in its infancy. To the best of our knowledge, no theoretical and computation study has been reported on the phase-separation of fluids with annealed disorder. The presence of impurity in the fluid systems plays an important role in domain growth and morphology which is a subject of interest in science and industry. These systems carry new perspectives pertaining to their experimental importance. 

In this paper, we undertake an extensive numerical study of domain growth dynamics of an immiscible symmetric binary fluid mixture in the presence of annealed disorder. The main aim of this work is to investigate the effect of disorder on the phase separation kinetics and aging phenomena of the fluid. The particular nature of the disorder used in our system is discussed in Section II.  We observe a dramatic slowing down in the growth dynamics of the domains in the presence of impurity particles. This is quantified in terms of the length scale $\ell(t)$, computed from the correlation function ${C}_{\psi\psi}(\vec{r},t)$. An algebraic domain growth is observed with the scaling exponent strongly dependent on disorder. We also explore the effect of impurities on the aging dynamics of our system by computing the two-time order parameter autocorrelation function $C_{ag}(t,t_w)$. The scaling laws of these correlation functions are verified for the disordered system.

The paper is organized as follows. In the next section, we outline the model and the numerical method to study the phase separation kinetics of segregating liquid mixture with annealed disorder. The results of the domain growth dynamics and the aging behavior are presented in Section III. Finally, in section IV, we offer a summary and a discussion of the results presented in this paper.
\section{Numerical Model and Method} 
\subsection{The basic model for binary liquid}
 For our present analysis, we perform a three-dimensional (3D) molecular dynamics (MD) simulation on a binary liquid in the NVT ensemble. We consider a $50:50$ mixture of A and B particles at high density $\rho= N/V=1$, where N and V represent the number of particles and volume of the system respectively. The two species interact via Lennard-Jones (LJ) potential
\begin{equation}
 U_{\alpha\beta}(\vec{r}) = 4\epsilon_{\alpha\beta}\Big[\Big(\frac{\sigma_{\alpha\beta}}{\vec{r}}\Big)^{12} - \Big(\frac{\sigma_{\alpha\beta}}{\vec{r}}\Big)^{6} \Big] 
 \end{equation}
  where $\ \vec{r}=|\vec{r_i}-\vec{r_j}|$ and $\alpha, \beta \in  \rm{A, B}$. To ensure energetically favorable phase separation, the parameters in the LJ potential are chosen as follows: $\sigma_{AA} = \sigma_{BB} = \sigma_{AB} = 1.0$ and $\epsilon_{AA} = \epsilon_{BB} = 1.0, \epsilon_{AB} = 0.5$. The choice of our interaction strength corresponds to the critical temperature $T_c=1.42$, outlying the possible liquid-solid and gas-liquid transition point \cite{Das}. The temperature is measured in units of $\epsilon/k_B$, where $k_B$ is Boltzmann’s
constant. Length is measured in units of $\sigma$. For simplicity we set the mass $m_0$ of A and B particles and $k_B$ equal to unity. For the sake of computational efficiency, the interaction potential is truncated to zero at $r_c=2.5\sigma$. Periodic boundary condition is applied in all three directions.
\subsection{Modeling annealed disorder}
To incorporate annealed disorder in our system, we choose a small fraction of  A-type particles and replace them with
 marked particles denoted by P. Therefore, the mixture comprises of A, B and P type of particles, all interacting via LJ
  potential.  We stress here that we are not trying to replicate a particular experimental system, but rather understand
   the effect of a particular choice of annealed disorder on phase separation kinetics. Of course, we can choose the 
   impurity particles in many ways. For example the choice of the energy parameters 
   $\epsilon_{PP},\epsilon_{PA},\epsilon_{PB}$ could be different, as well as the corresponding interaction length 
   scales. Other possibilities could be multi-body interactions, angle-dependent interactions, etc. For concreteness, 
   in our present analysis, we choose the same energy or length-scale parameters for P as the A particles. The only 
   change made is the mass of the P particles that reduces their mobility. The presence of the P particles acts as an 
   impediment and influences the dynamics of the whole system. 

 The interaction parameters between the P particles and with the rest of the system are as follows: $\sigma_{PP}=\sigma_{PA}=\sigma_{PB}=1.0$, $\epsilon_{PP}=1.0$, and $\epsilon_{PA}=\epsilon_{PB}=0.5$. The interaction potential for the P particles is truncated at $r_c=2^{\frac{1}{6}}\sigma$ to exclude any attractive force and avoid clustering. The mobility of the impurity is varied by changing their mass $m_P$. The simulation is performed with five different choice of masses, $m_P=1,10000,100000$ and $\infty$. The two limits $m_P=1$ and $m_P=\infty$ correspond to the pure system free from any disorder and the system with immobile quenched disorder. 

  We performed the MD simulations using the Velocity Verlet algorithm \cite{Verlet} over a total of 262144 ($=64^3$) particles for two different impurity concentrations. The density of the impurity particles is chosen such that $\rho_{\rm{P}} \ll \rho_0$ where $\rho_{\rm{P}}$ and $\rho_0$ are the densities of the impurity and the rest of the particles respectively with $\rho=\rho_{\rm{P}}+\rho_0$. The mass $m_P$ is varied with the choice mentioned above. The $m_P=\infty$ case is replicated by not allowing the P particles to participate in the dynamics and stay frozen in their respective positions. We begin our simulation by preparing a well equilibrated homogeneous mixture at high temperature $T=10.0$ followed by a quench to $T=0.77 T_c$ at $t=0$. Here the reduced unit of time is taken as $\sigma\sqrt{m/\epsilon}$. Temperature is controlled by Nose-Hoover thermostat (NHT) which preserves the hydrodynamic effect \cite{Nose}. Finally, the system is allowed to evolve to the thermodynamically favored state until the complete phase separation is achieved. The ensemble average of all the statistical quantities is obtained from 10 independent runs at $0.77T_c$ starting from completely different initial configurations. \\
\subsection{Correlation function and Structure factor}
To characterize the domain morphology and study the domain growth we introduce the two point equal time correlation function $C_{\psi\psi}(\vec{r},t)$ as follows:
\begin{equation}\label{Correlation_function}
C_{\psi\psi}(\vec{r},t) = \langle\psi(0,t)\psi(\vec{r},t)\rangle /\langle\psi(0,t)\rangle^2
\end{equation} 
where the order parameter $\psi(\vec{r},t)$ is obtained as follows: we compute the local density difference $\delta\rho = \rho_A-\rho_B$ between the two species A and B, calculated over a box of size $(2\sigma)^3$ located at $\vec{r}$. The $\psi(\vec{r},t)$ is assigned a value $+1$ when $\delta\rho>0$, and $-1$ otherwise. The angular brackets indicate the statistical averaging. We  also compute the structure factor $S(\vec{k},t)$ by taking the Fourier transform of the correlation function given by 
\begin{equation}\label{Structure_factor}
S(\vec{k},t) = \int d\vec{r} \hspace{0.1cm} exp(i\vec{k}.\vec{r})  \hspace{0.1cm} C_{\psi\psi}(\vec{r},t)
\end{equation}
Finally for the isotropic system, spherically averaged $C_{\psi\psi}(r,t)$ and $S(k,t)$ are computed.

To study the aging dynamics we considers the so-called two-time order-parameter correlation function $C_{ag}(t,t_w)$ as follows:
\begin{equation}\label{autoCorrelation_function}
C_{ag}(t,t_w) = \langle\psi(\vec{r},t)\psi(\vec{r},t_w)\rangle - \langle\psi(\vec{r},t)\rangle\langle\psi(\vec{r},t_w)\rangle
\end{equation}
where $t$ is the observation time and $t_w$ is the waiting time or the age of the system after quench.
\section{Numerical Results} 
\subsection{Domain morphology and growth dynamics}
\begin{figure}[h!]
	\centering
\includegraphics[width=\columnwidth]{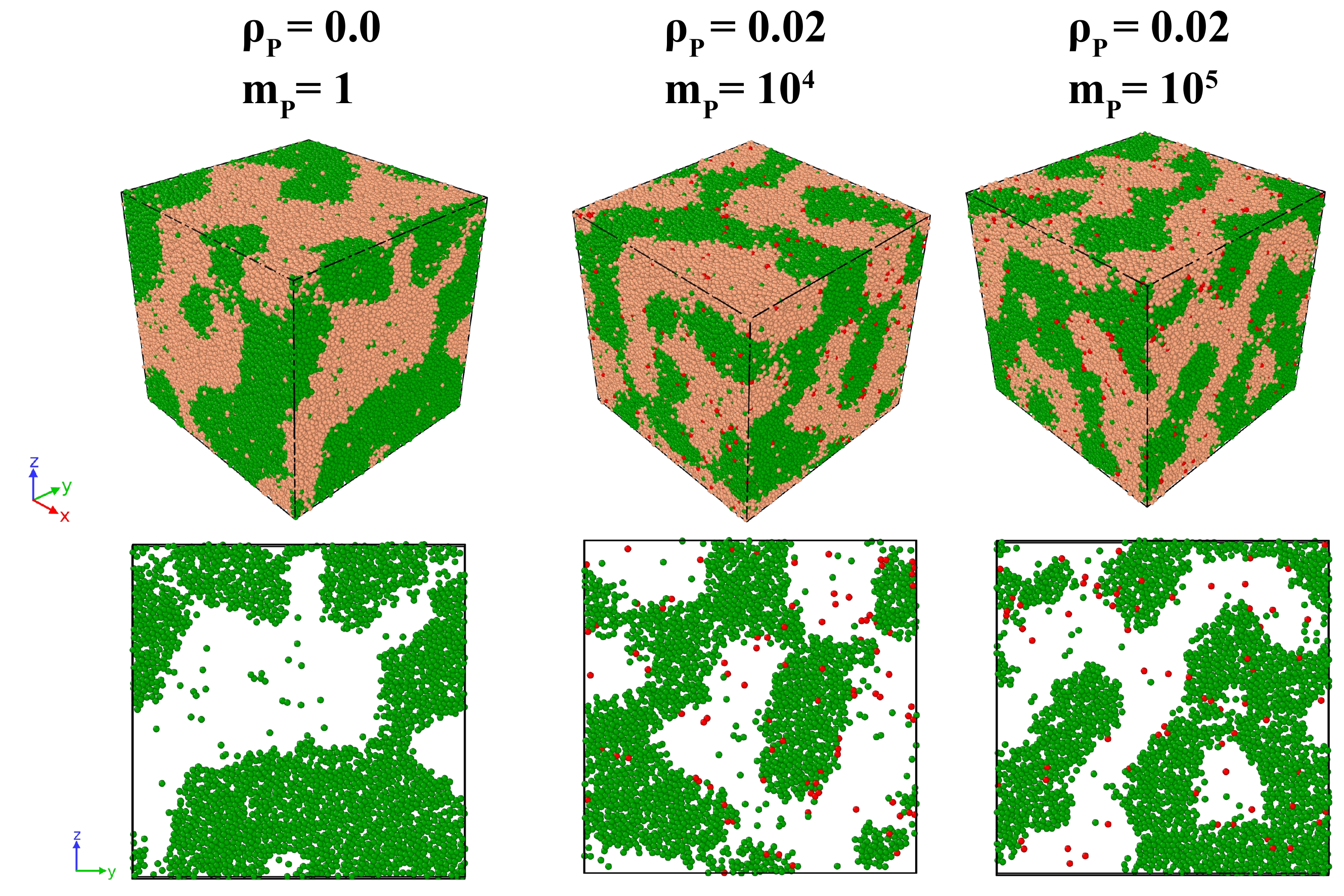}
\caption{Typical snapshots of our phase separating system for $\rho_{\rm{P}}=$ 0.0 and 0.02 at time 
$t=2000$ (LJ unit) are shown in the upper panel. The A, B and P particles are marked as orange, green, and red respectively. A two dimensional cross sectional area of the configurations with B particles is shown in the lower panel.}
\label{snap2}
\end{figure}

The effect of annealed disorder on the phase separation dynamics is depicted in Fig.~\ref{snap2}. Here we show the 3-dimensional configurations obtained from our simulation for three different $m_p$ at the impurity concentration $\rho_{\rm{P}}=0.02$ at time $t=2000$. We observe a bicontinuous A-rich and B-rich domains. It is apparent from the snapshots that the same species cluster sizes get smaller with increase in $m_P$ at a given time. Therefore, the kinetics of ordering slows down with decrease in mobility of the impurity particles. The same effect is observed with disorder concentration $\rho_{\rm{P}}=0.01$ also. This will be quantified in detail in terms of the domain length-scale $\ell(t)$ later. 

The slowing down of kinetics can be contemplated in the following way. In the course of phase separation, the same species domains form and grow with time. Careful visual inspections of our simulation data reveal that majority of the impurity P type particles are located across these domain boundaries all the time. This is evident from the 2D cross-sectional plot of the configurations shown in Fig.~\ref{snap2}. As a result, they localize the domain walls in energetically favorable positions. However, this is overcome by the thermal energy due to sufficiently high temperature. It is worth noting in Fig.~\ref{snap2} that the domain boundaries become rough with impurity. The implications of the modified domain boundary will be discussed shortly in terms of the correlation function and structure factor. 
\par
\begin{figure}[h!]
	\centering
	\includegraphics[width=\columnwidth]{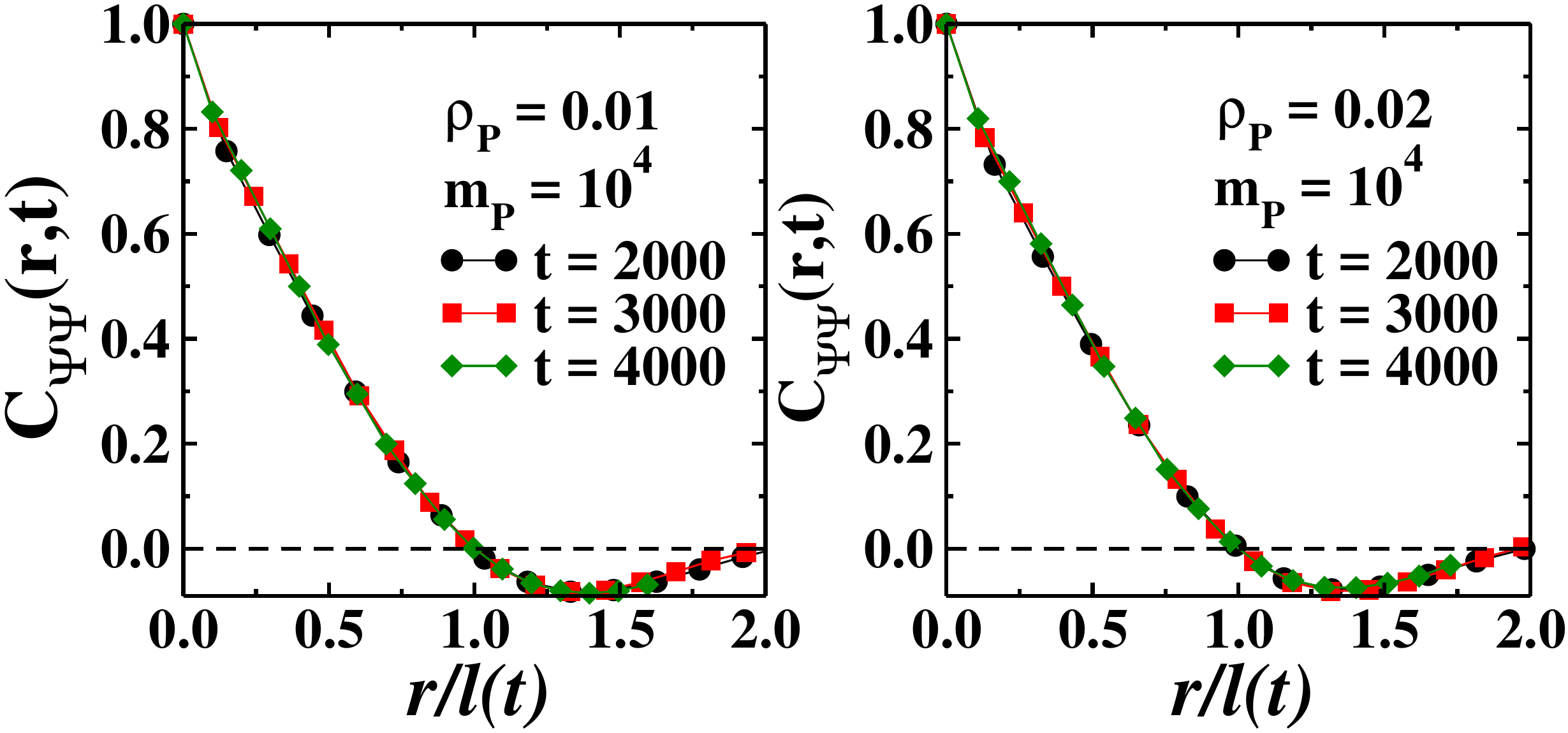}
	\caption{The scaling plots of $C_{\psi\psi}(r,t)$ vs $r/\ell(t)$ are shown for $\rho_{\rm{P}}= 0.01$ and 0.02 at $m_P=10000$.}
	\label{fig1}
\end{figure}

\begin{figure}[h!]
	\centering
	\includegraphics[width=\columnwidth]{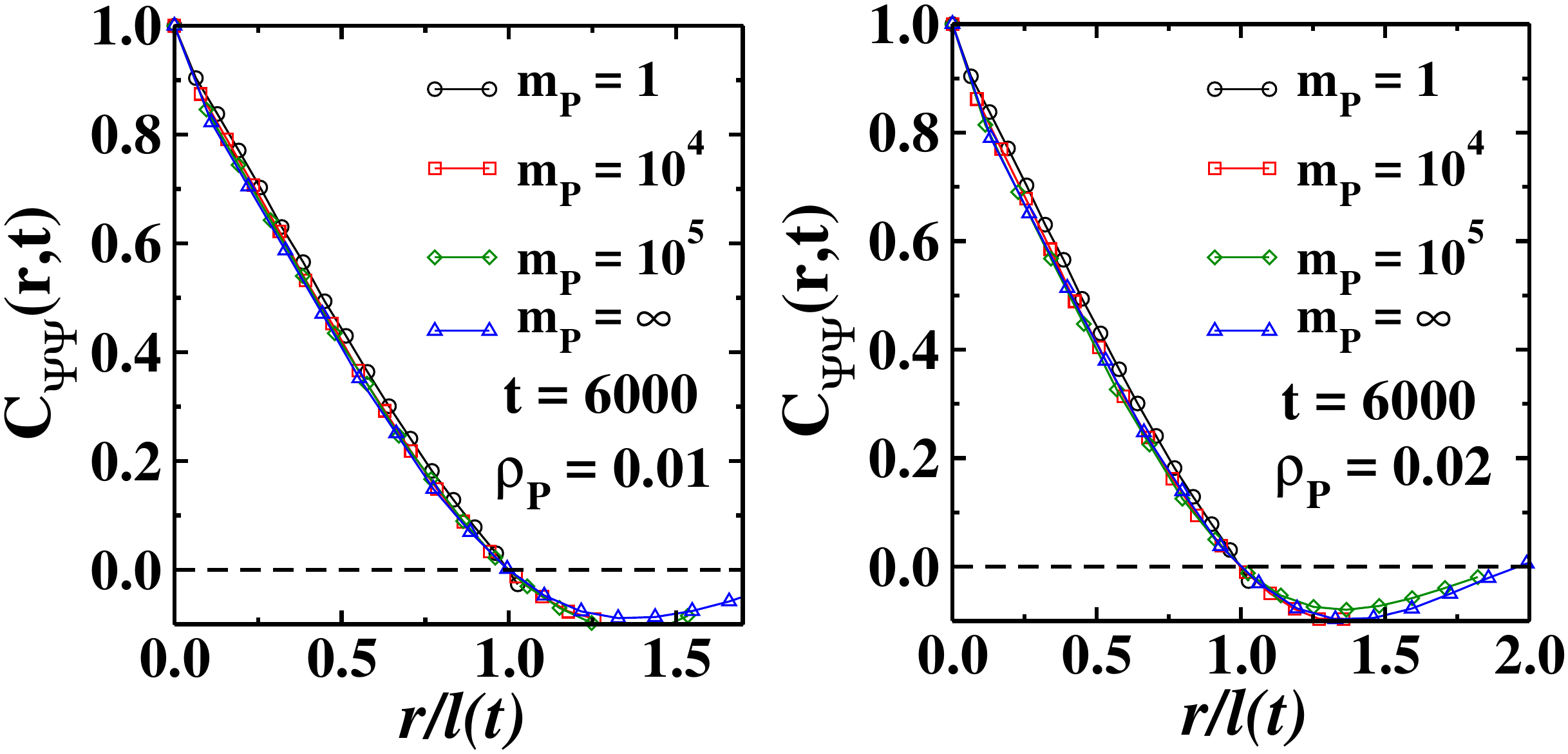}
	\caption{Scaled correlation function $C_{\psi\psi}$ vs $r/\ell$ plot for $\rho_{\rm{P}}=0.01$ and 0.02 at time $t=6000$ for different $m_P$.}
	\label{fig2}
\end{figure}

 To gain a qualitative understanding of the domain evolution for disordered system we compute the correlation function $C_{\psi\psi}(r,t)$ given by Eq. 2. In Fig.~\ref{fig1} we show the scaling plot for $C_{\psi\psi}(r,t)$ vs $r/ \ell(t)$ for $\rho_{\rm{P}}=0.01$ and 0.02 at $m_P=10000$, where $\ell(t)$ is the average domain size. There are quite a few ways to quantify the $\ell(t)$, e.g, first zero crossing of $C_{\psi\psi}(r,t)$, half-crossing of $C_{\psi\psi}(r,t)$, inverse of the first moment of $S(k,t)$. We have used the first method to calculate $\ell(t)$. An excellent data collapse is observed for different times. We repeated the same exercise with other values of $m_P$ and found the quality of overlap to be the same.  
This suggests that even in the presence of disorder (fixed $\rho_{\rm{P}}$), the system belongs to the same dynamical universality class \cite{Binder3}. However, the correlation functions corresponding to different $m_P$ for a given impurity concentration do not overlap with each other when plotted at a fixed time $t$. This is shown in Fig.~\ref{fig2}. For the pure system the $C_{\psi\psi}(r,t)$ shows a linear behavior at small $r$ following the Porod law, $C(r,t) \sim 1 - ar$. However, in the presence of impurity, a nonlinear or cusp nature is observed, which can be attributed to the scattering from the rough domain boundary \cite{Gaurav}. Therefore, the annealed disorder results in breaking down the Porod law in the correlation function \cite{Shaista}. 
\par
\begin{figure}
	\centering
	\includegraphics[width=0.8\columnwidth]{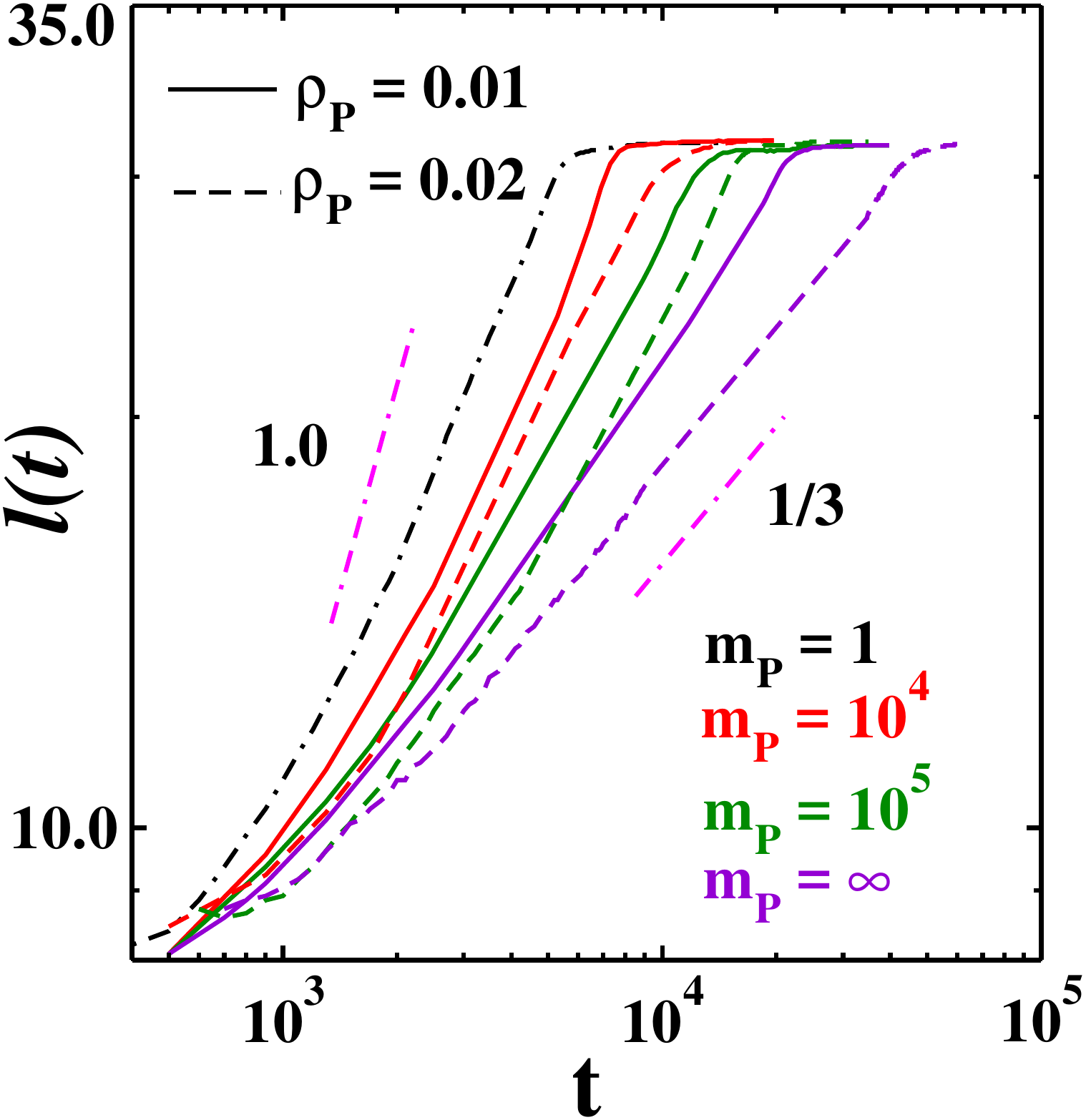}
\caption{(a) The time evolution of the average domain size $\ell(t)$ for different masses with $\rho_{\rm{P}} = 0.01$ (solid curve) and 0.02 (dashed curve) in the log-log scale. The dashed dotted line represents the $\ell(t)$ for the pure system. The double-dashed dotted lines represent the guideline for the slope.}
\label{fig3}
\end{figure}

 The evolution morphology is best realized in terms of the domain size $\ell(t)$. In Fig.~\ref{fig3} we show the time dependence of $\ell(t)$ for different impurity concentrations. Our simulation is able to access the viscous hydrodynamic regime after an initial transient period. For the pure system ($\rho_{\rm{P}}=0$) we expect here a linear behavior of $\ell(t)$. The slight deviation from this can be attributed to the nonzero off-sets at the crossovers which can be subtracted from $\ell(t)$ to recover the proper linear behavior \cite{skd}. In the presence of impurity the growth rate slows down and the total time taken for the system to be completely phase-separated increases significantly with disorder. This is consistent with the observations in Fig.~\ref{snap2}. The slowest growth for the system under consideration ($\rho_{\rm{P}}=0.02$ and $m_P=\infty$) corresponds to $\alpha \sim 1/3$.\\
In the presence of disorder, the relaxation mechanism changes from pure energy lowering to thermally activated process, due to the creation of energy barriers. The presence of the energy barrier impedes the relaxation process. An important issue in this regard is the so-called superuniversality (SU), i.e,  whether the disorder-dependent spatial autocorrelation function scales to a master curve when the length is rescaled with respect to the domain size $\ell(t)$ \cite{Lai,Corberi}. From Fig.~\ref{fig2} it is conspicuous that the disordered system does not fall into the SU class. We also observe an algebraic domain growth throughout the phase separation process with a disorder-dependent exponent. This implies that the barrier size does not depend on the domain size for a given $\rho_{\rm{P}}$. \\
\begin{figure}
	\centering
	\begin{subfigure}[h!]{0.5\columnwidth}
	\includegraphics[width=\columnwidth]{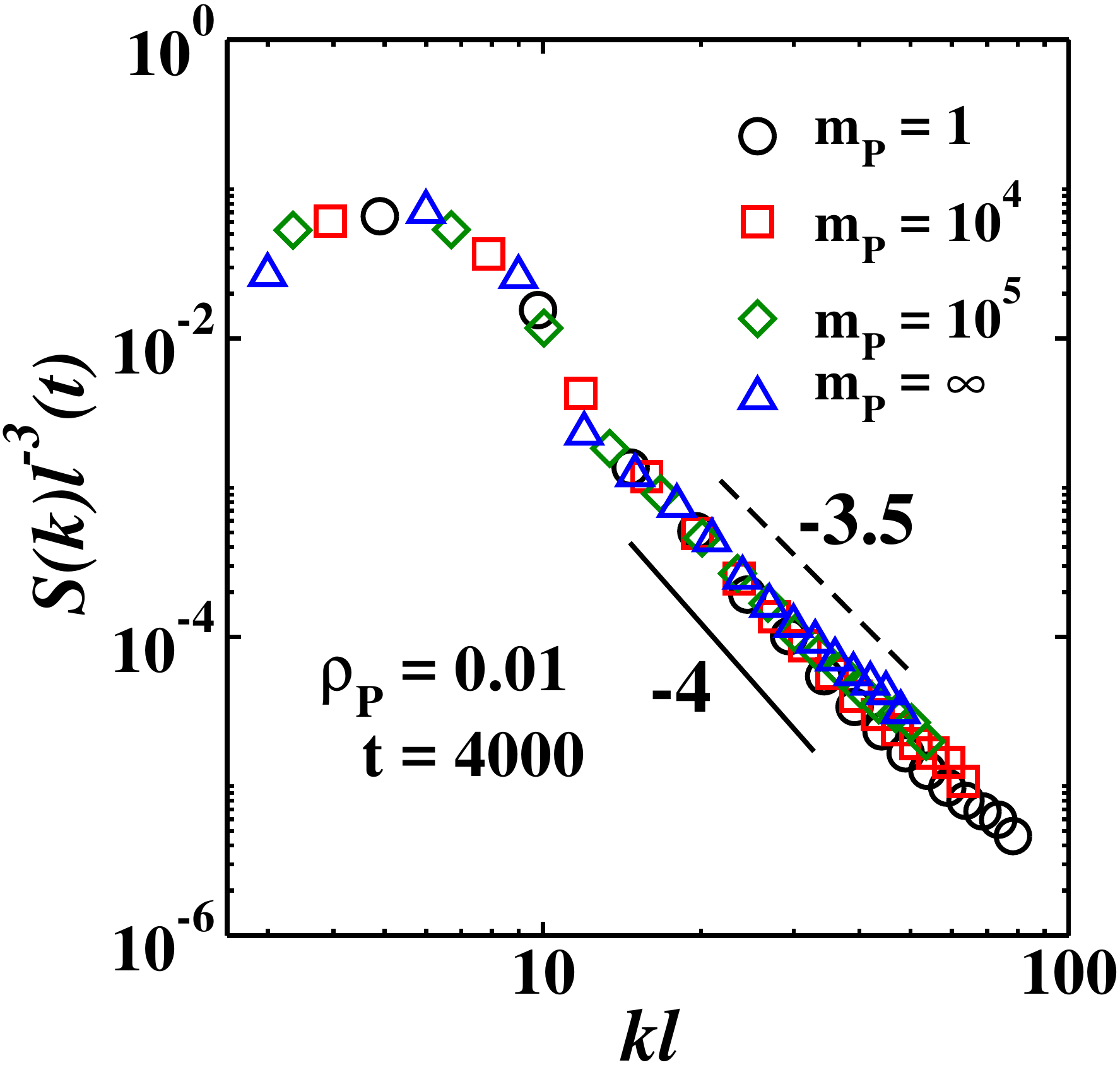}
\end{subfigure}%
\begin{subfigure}[h!]{.5\columnwidth}
	\centering
	\includegraphics[width=\columnwidth]{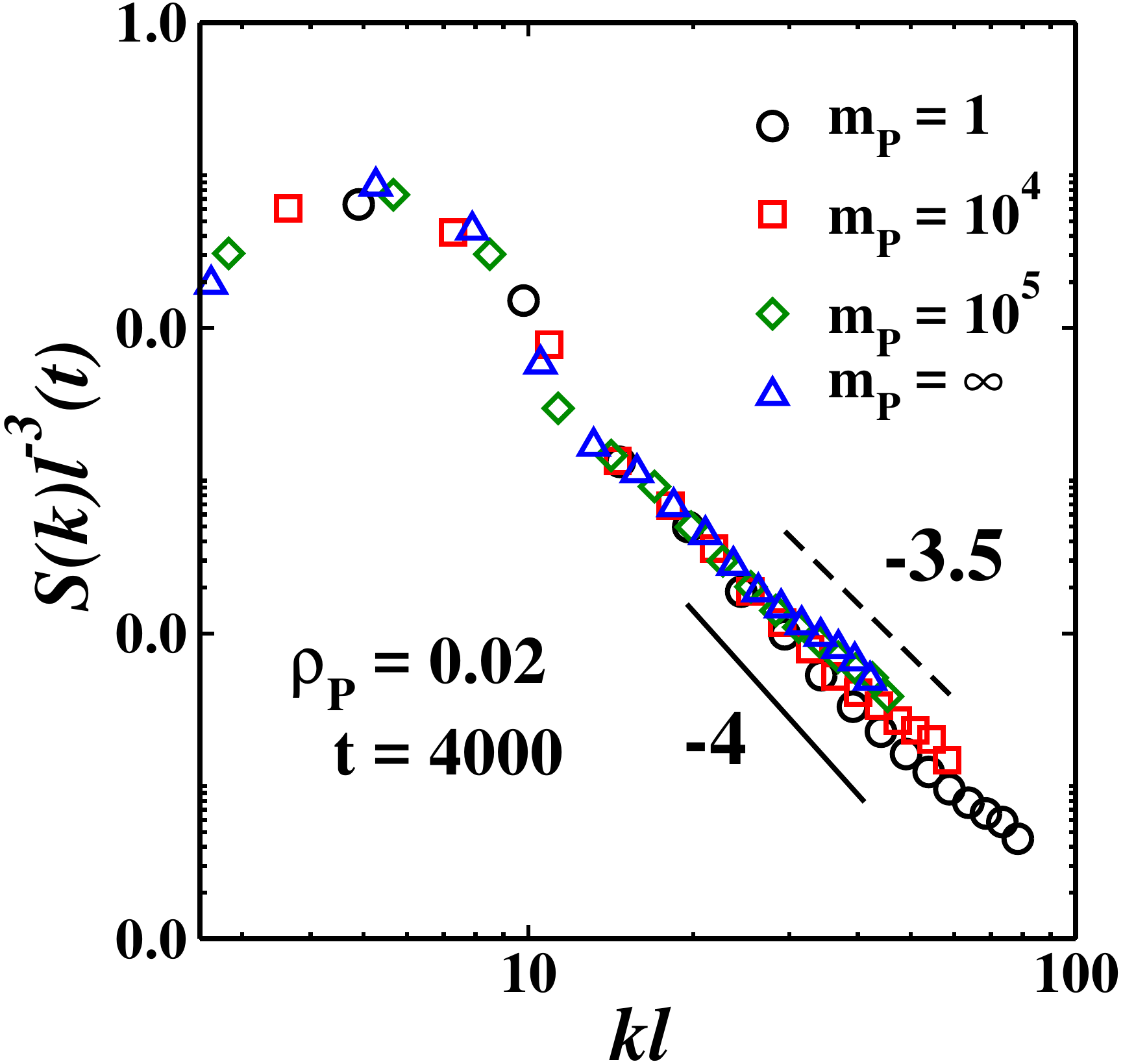}
\end{subfigure}
\caption{(a) The scaled structure factor $S(k)\ell^{-3}$ vs $\ell k$ is plotted (points) for various $m_P$ with $\rho_{\rm{P}}=$ 0.01 and 0.02 in the log-log scale. The straight line is guideline for the Porod law $S(k) \sim k^{-4}$ and the dashed line for non-Porod behavior $S(k) \sim k^{-3.5}$ (see text).}
\label{SK}
\end{figure}
In Fig.~\ref{SK}, we show the scaled structure factor $S(k,t) \ell^{-3}$ vs $\ell k$ plot for different $\rho_{\rm{P}}$ and $m_P$ in log-log scale. Although, the decaying part of the tail for the pure system shows the Porod law behavior $S(k,t) \sim k^{-(d+1)}$ \cite{Binder3}, the disordered systems show a non-Porod behavior, $S(k,t) \sim k^{-(d+1+\theta)}$, where $\theta \simeq -0.5$. This value is very close to the results observed in the 3D random field Ising model \cite{Gaurav}. This substantiates the roughening of the domain interfaces with annealed disorder and the violation of SU.
\begin{figure}
	\centering
	\includegraphics[width=0.8\columnwidth]{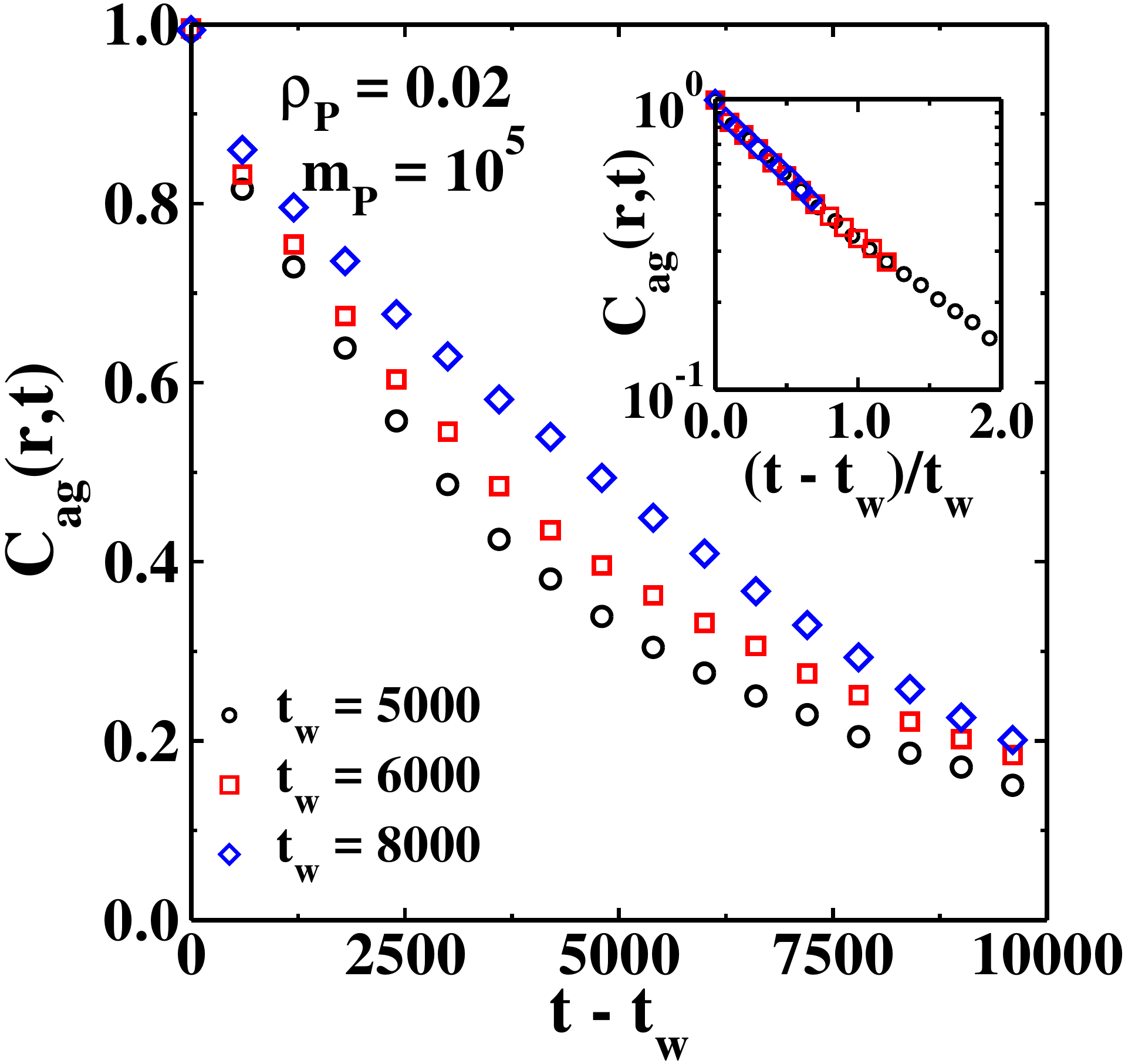}	
\caption{ The two point order parameter autocorrelation function $C_{ag}(t,t_w)$ vs $t-t_w$ plot for $\rho_{\rm{P}}=0.02$ and $m_P=10^5$ at three different waiting time $t_w=$ 5000, 6000, and 7000. In the inset we show the same data after rescaling the time with $t_w$.}
\label{fig9}
\end{figure}
\subsection{Aging dynamics}
\par Now we turn our focus to one of the most important aspects of non-equilibrium dynamics namely the aging phenomena. As discussed in the introduction, this subject is well studied for the disorder-free binary mixture systems. Therefore, we abstain from showing results on pure mixtures. In Fig.~\ref{fig9} we show the variation of $C_{ag}(t,t_w)$ with $t - t_w$ for the disordered system with maximum chosen impurity. For our aging related studies, we always focus on the viscous hydrodynamics regime where the effect of impurities is prominent (as reflected in Fig. \ref{fig3}) and the $t_w$ are chosen accordingly. Clearly, the correlation curves corresponding to different $t_w$ do not overlap, demonstrating the violation of time translation invariance. Following Fisher and Huse \cite{Fisher1} we attempt to scale the abscissa as $t/t_w$ and a nice data collapse is obtained as shown in the inset of Fig.~\ref{fig9}. Therefore, the Fisher and Huse scaling law remains vindicated for the systems with annealed disorder.  
\begin{figure}
	\centering
	\includegraphics[width=0.8\columnwidth]{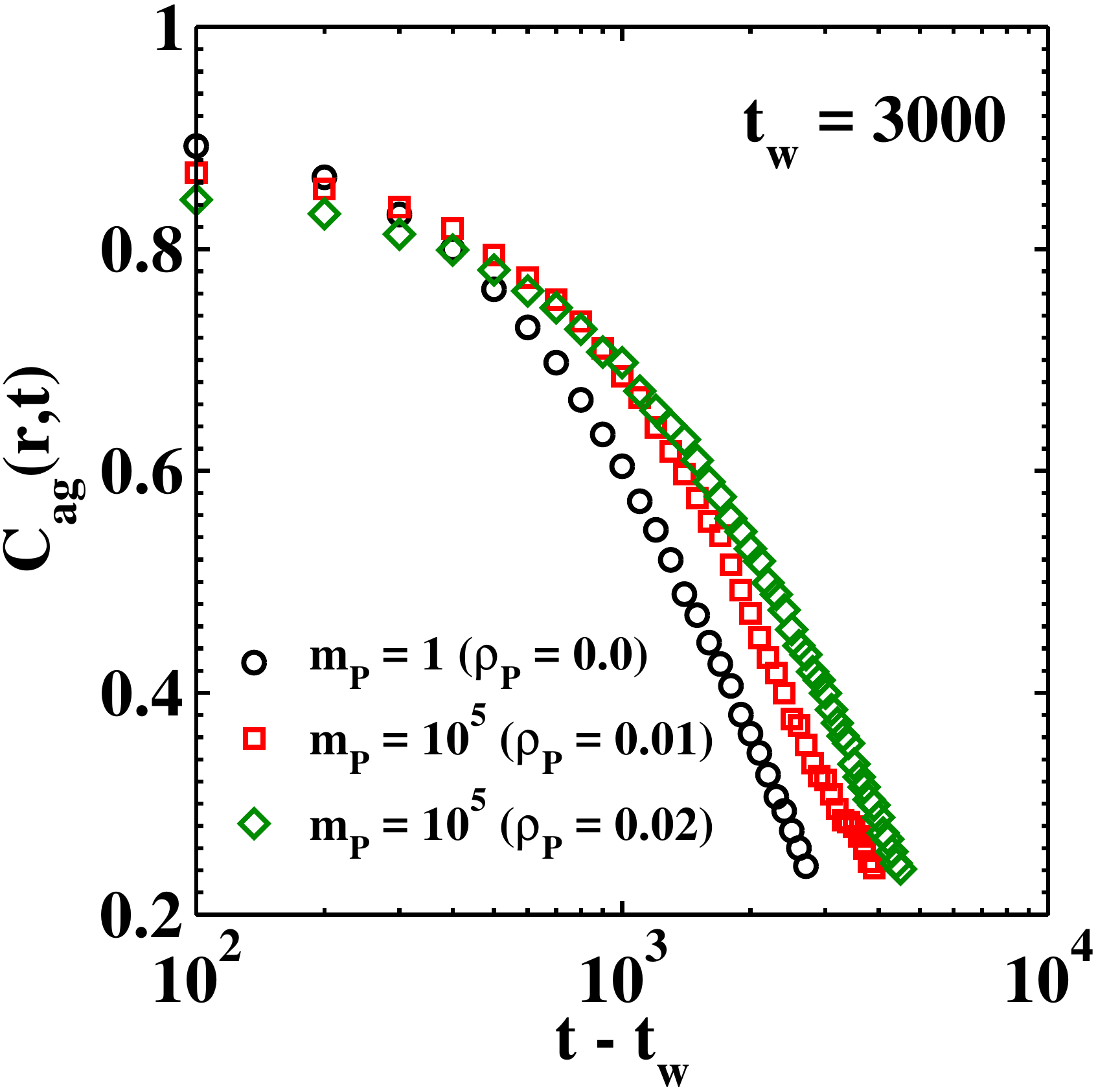}
	\caption{The plot of $C_{ag}(t,t_w)$ vs $t-t_w$ for the pure ($\rho_{\rm{P}}=0.0$) and impure ($\rho_{\rm{P}}=$ 0.01 and 0.02; $m_P=10^5$) system at $t_w=3000$. }
	\label{fig6}
\end{figure}
\par The role of impurity in the aging dynamics is further investigated by computing the $C_{ag}(t,t_w)$ for the pure and disordered systems at a fixed $t_w=3000$. This is shown in Fig.~\ref{fig6}. Clearly, the correlation function decays significantly slowly with increase in impurity concentration. We checked and found similar behavior when the heaviness $m_P$ of the impurity particles is changed to other values also. This demonstrates the slowing down of the dynamics and is consistent with the observation in Fig.~\ref{fig3} .

\begin{figure}
	\begin{subfigure}[vt!]{0.63\linewidth}
	\centering
	\includegraphics[width=\columnwidth]{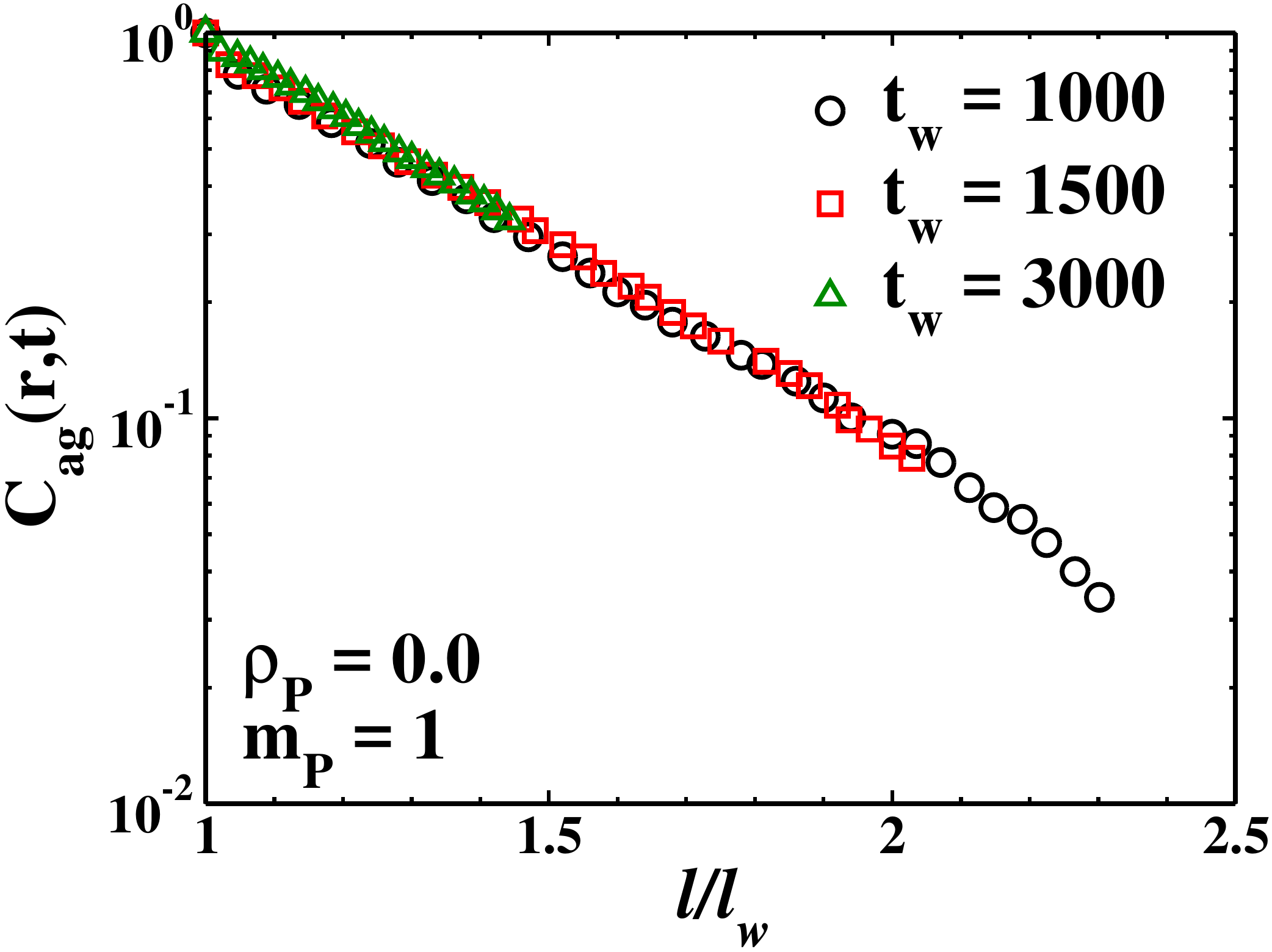}
\end{subfigure}
\begin{subfigure}[vt!]{.6\linewidth}
	\centering
	\includegraphics[width=\columnwidth]{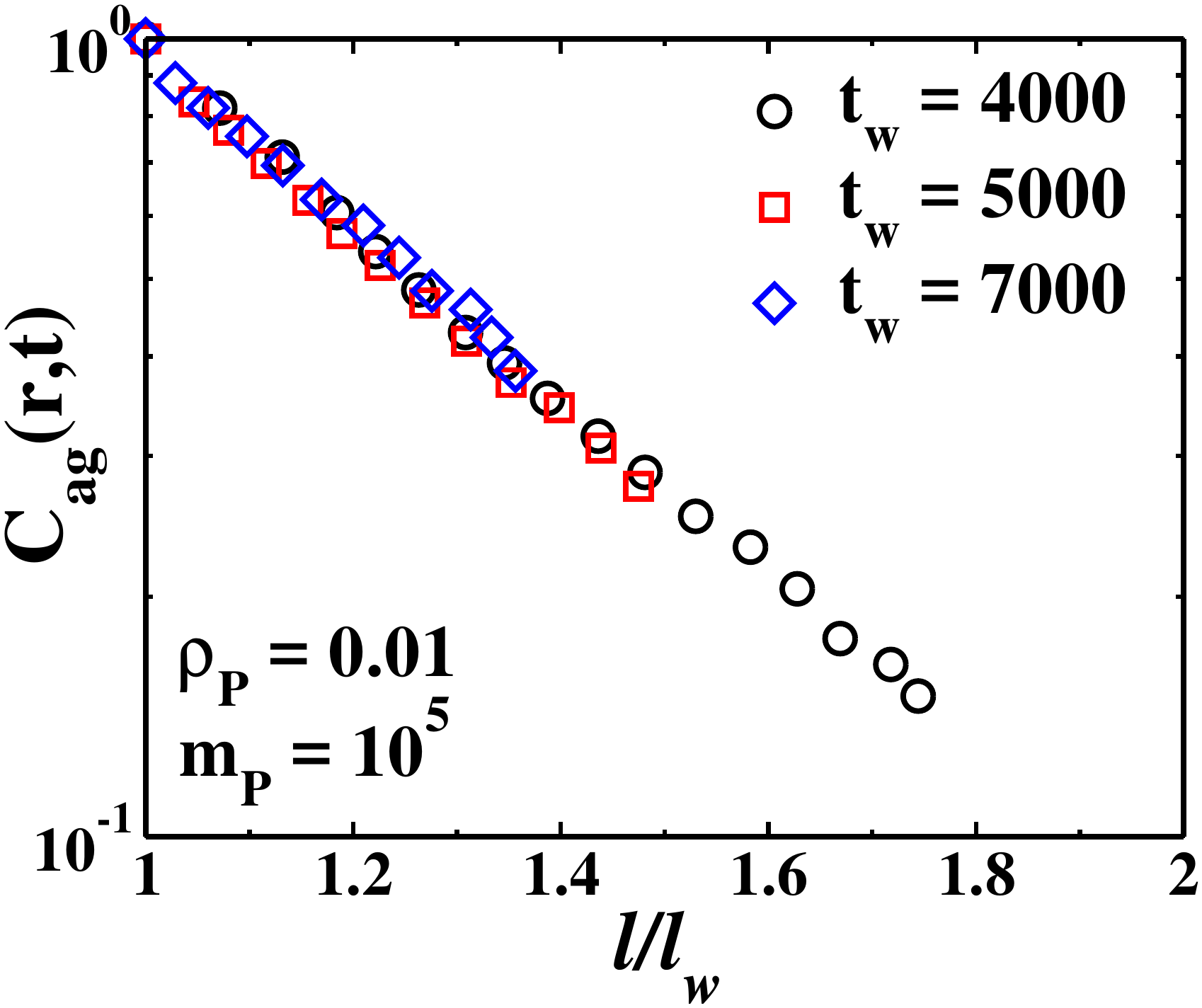}
\end{subfigure}
\begin{subfigure}[vt!]{.6\linewidth}
	\centering
	\includegraphics[width=\columnwidth]{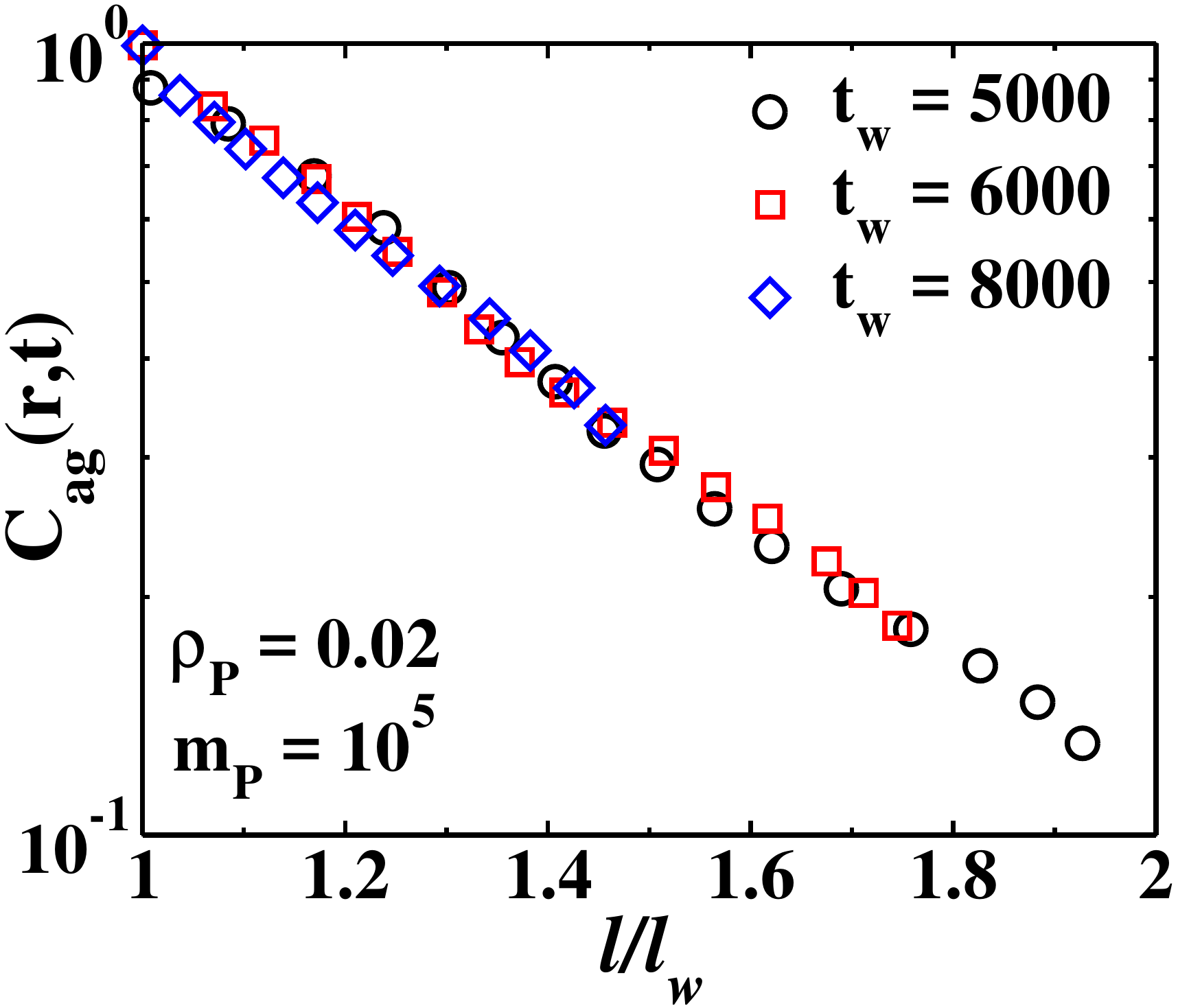}
\end{subfigure}
\caption{ The scaling of the correlation function $C_{ag}(\vec{r},t)$ plotted as a function of $l/l_w$ for the pure ($\rho_{\rm{P}}=0$) and the disordered ($\rho_{\rm{P}}=0.01, 0.02$) system for different values of $t_w$ in the semi-log scale.}
\label{fig7}
\end{figure}
\par In Fig.~\ref{fig7} we show the $C_{ag}(\vec{r},t)$ as a function of $l/l_w$. An excellent data collapse is found for the pure as well as the impure system with different degrees of disorder. To tally the nature of the master curve with the previously found exponential decay in the pure binary LJ system \cite{sahmed}, we plot the data on the semi-log scale. The data set appears to be linear confirming the exponential nature. Therefore, we find the scaling law in the hydrodynamic regime is very generic, insensitive to the annealed disorder. The exponential decay of the correlation function is attributed to the advective hydrodynamic flows under the hydrodynamic effect that causes large displacement of particles \cite{sahmed}. 

\section{Summary and Discussion}
In summary, we have examined the kinetics of phase separation in a binary liquid in the presence of annealed disorder. The disorder was introduced by adding a small concentration of foreign particles to the system. The key observation is the dramatic slowing down of the phase separation dynamics in the presence of the impurity particles. The spatial correlation function scaled with domain size showed disorder dependence and non-Porod behavior which can be attributed to the roughening of domain boundaries. The pure system clearly exhibits power-law domain growth with exponent unity in the viscous hydrodynamic regime. In the presence of disorder, the nature of growth remains algebraic but the exponent decreases with impurity concentration and its mobility. The disorder localizes the domain wall and creates an energy barrier that is overcome by the thermal activation process. However, the algebraic domain growth conforms to the domain size-independent energy barrier throughout the phase separation process. We have also examined the structure factor for our system. The decaying part of the structure factor gets modified in the presence of disorder and shows a non-Porod tail. Both the spatial correlation function and the structure factor do not obey superuniversality. \\
We have also investigated the effect of annealed disorder on aging phenomena. The aging dynamics is quantified by computing the two-time order-parameter autocorrelation function $C_{ag}(t,t_w)$. We demonstrated that $C_{ag}(t,t_w)$ follows the scaling law with respect to $t/t_w$ proposed by Fisher and Huse in the disordered system. But the aging process slows down significantly as reflected in the decay of $C_{ag}(t,t_w)$ with time. We have also examined the scaling law with respect to $\ell/\ell_w$ and obtained an excellent data collapse for all the cases in hand. The scaling function showed an exponential decay in the hydrodynamic regime. We, therefore, conclude that the scaling laws are very robust and generic for the aging dynamics and holds equally good for the disordered system. We believe that our model and the results presented here offer a method to control the phase separation dynamics in real systems by adding annealed disorder appropriately. Studies following this line of thought will be presented elsewhere. \\

\noindent{\it Acknowledgment.---} B.S.G. acknowledges Science and Engineering Research Board (SERB), Dept. of Scien ce and Technology (DST), Govt. of India (no. SRG/2019/001923) for financial support.


\begin{thebibliography}{99}
	
	\bibitem{Fisher}
	M.E. Fisher, Rep. Prog. Phys. \textbf{30}, 615–730 (1967).
	\bibitem{Stanley}
	H.E. Stanley, Introduction to Phase Transitions and Critical Phenomena (Oxford University Press, 1971).
	\bibitem{Binder}
	 K. Binder, Rep. Prog. Phys. \textbf{50}, 783–859 (1987).
	 \bibitem{Jones}
	 R.A.L. Jones, Soft Condensed Matter (Oxford University Press, 2002).
	 \bibitem{Bray}
	 A.J. Bray, Adv. Phys. \textbf{51}, 481–587 (2002).		
	 \bibitem{Binder1}
	 K. Binder, Phys. Rev. B \textbf{15}, 4425–4447 (1977).
	 \bibitem{Siggia}
	 E. D. Siggia, Phys. Rev. A \textbf{20}, 595 (1979).
	 \bibitem{Furukawa}
	 H. Furukawa, Phys. Rev. A \textbf{31}, 1103–1108 (1985).
	 \bibitem{Miguel}
	 M. San Miguel, Phys. Rev. A \textbf{31}, 1001–1005 (1985).
	 \bibitem{Tanaka}
	 H. Tanaka, J. Chem. Phys. \textbf{103}, 2361 (1995).
	 \bibitem{Beysens}
	  D. Beysens, Y. Garrabos, D. Chatain, Europhys. Lett. \textbf{86}, 16003 (2009).
	  \bibitem{Tanaka1}
	 S. Tanaka, Y. Kubo, Y. Yokoyama, A. Toda, K. Taguchi, H. Kajioka, J. Chem. Phys. \textbf{135}, 234503 (2011).
	 \bibitem{Kendon}
	 V.M. Kendon, M.E. Cates, I. Pagonabarraga, J.C. Desplat, P. Blandon, J. Fluid Mech. \textbf{440}, 147–203 (2001).
	 \bibitem{Puri}
	 S. Puri, B. Dünweg, Phys. Rev. A \textbf{45}, R6977–R6980 (1992).
	 \bibitem{Dutt}
	 C. Datt, S.P. Thampi, R. Govindarajan, Phys. Rev. E \textbf{91}, 010101(R) (2015).
	 \bibitem{Laradji}
	 M. Laradji, S. Toxvaerd, O.G. Mountain, Phys. Rev. Lett. \textbf{77}, 2253–2256 (1996).
	 \bibitem{Thakre}
	 A.K. Thakre, W.K. den Ohe, W.J. Briels, Phys. Rev. E \textbf{77}, 011503 (2008).
	 \bibitem{Ahmad}
	 S. Ahmad, S.K. Das, S. Puri, Phys. Rev. E \textbf{82},  040107 (2010).
	 \bibitem{Binder2}
	 K. Binder and D. Stauffer, Phys. Rev. Lett. \textbf{33}, 1006 (1974); Z. Phys. B \textbf{24}, 407 (1976).
	 \bibitem{Binder3}
	 K. Binder, in Phase Transformation of Materials, edited by R. W. Cahn, P. Haasen, and E. J. Kramer, Material Science and Technology (VCH, Weinheim, 1991), Vol. 5, p. 405; Kinetics of Phase Transitions, edited by S. Puri and V. Wadhawan (CRC Press, Boca Raton, 2009).
	 \bibitem{Fisher1}
	 D.S. Fisher and D.A. Huse, Phys. Rev. B \textbf{38}, 373 (1988).
	 \bibitem{Yeung}
     C. Yeung, M. Rao, and R.C. Desai, Phys. Rev. E 53, 3073 (1996).
     \bibitem{sahmed}
     S. Ahmad, F. Corberi, S. K. Das, E. Lippiello, S. Puri, and M. Zannetti, Phys. Rev. E \textbf{86}, 061129 (2012).
\bibitem{Puri1}
S. Puri, Phase Transitions \textbf{77}, 469 (2004).
\bibitem{Huse}
D. A. Huse and C. L. Henley, Phys. Rev. Lett. \textbf{54}, 2708 (1985).
\bibitem{Grest}
G. S. Grest and D. J. Srolovitz, Phys. Rev. B \textbf{32}, 3014 (1985).
\bibitem{Srolovitz}
D. J. Srolovitz and G. S. Grest, ibid. \textbf{32}, 3021 (1985).
\bibitem{Puri2}
S. Puri, D. Chowdhury and N. Parekh, J. Phys. A \textbf{24}, L1087 (1991).
\bibitem{Bray2}
 A. J. Bray and K. Humayun, J. Phys. A \textbf{24}, L1185 (1991).
\bibitem{Rao}
M. Rao and A. Chakrabarti, Phys. Rev. Lett. \textbf{71}, 3501 (1993).
\bibitem{Paul}
R. Paul, S. Puri and H. Rieger, Europhys. Lett. \textbf{68}, 881 (2004); Phys. Rev. E \textbf{71}, 061109 (2005).
\bibitem{Henkel}
M. Henkel and M. Pleimling, Europhys. Lett. \textbf{76}, 561 (2006); Phys. Rev. B \textbf{78}, 224419 (2008).
\bibitem{Aron}
C. Aron, C. Chamon, L. F. Cugliandolo and M. Picco, J. Stat. Mech., P05016 (2008); L. F. Cugliandolo, Physica A \textbf{389}, 4360 (2010).
\bibitem{Brochard}
F. Brochard and P. G. de Gennes, J. Phys. Lett. (Paris) \textbf{44}, 785 (1983).
\bibitem{Gennes1}
P. G. de Gennes, J. Phys. Chem. \textbf{88}, 6469 (1984).
\bibitem{Maher}
J. V. Maher, W. I. Goldburg, D. W. Pohl and M. Lanz, Phys. Rev. Lett. \textbf{53}, 60 (1984).
\bibitem{Goh}
M. C. Goh, W. I. Goldburg and C. M. Knobler, Phys. Rev. Lett. \textbf{58}, 1008 (1987).
\bibitem{Wiltzius}
P. Wiltzius, S. B. Dierker and B. S. Dennis, Phys. Rev. Lett. \textbf{62}, 804 (1989).
\bibitem{Das}
S. K. Das, M. E. Fisher, J. V. Sengers, J. Horbach and K. Binder, Phys. Rev. Lett. \textbf{97}, 025702 (2006).
\bibitem{Verlet}
L. Verlet, Phys. Rev. \textbf{98}, 159 (1967).
\bibitem{Nose}
D. Frenkel and B. Smit, Understanding Molecular Simulations: From Algorithms to Applications (Academic Press, San Diego,
2002).
\bibitem{Gaurav}
G. P. Shrivastav, S. Krishnamoorthy, V. Banerjee and S. Puri, Europhys Lett. \textbf{96}, 36003 (2011).
\bibitem{Shaista}
S. Ahmad, S. Puri and S. K. Das, Phys. Rev. E \textbf{90}, 040302(R) (2014).
\bibitem{skd}
S.K. Das, S. Roy, S. Majumdar, and S. Ahmad, Europhys. Lett. \textbf{97}, 66006 (2012).
\bibitem{Lai}
Z. W. Lai, G. F. Mazenko and O. T. Valls, Phys. Rev. B \textbf{37}, 9481 (1988).
\bibitem{Corberi}
F. Corberi, E. Lippiello, A. Mukherjee, S. Puri and M. Zannetti, Phys. Rev. E \textbf{85}, 021141 (2012).

\end{thebibliography}
\end{document}